\documentclass[oneside,a4paper,reqno]{amsart} 
\usepackage{amsmath}
\usepackage{amssymb}
\usepackage{xspace}
\usepackage[mathscr]{eucal}
\usepackage{graphics}

\theoremstyle{remark}

\newcommand{\mat}[3]{\ensuremath{
																								\left \langle  \vphantom{#2 #3}   #1   
                        \right|    					\, #2\,   
                        \left|    \vphantom{#2 #1} #3   
                        \right \rangle
                        								}
                     }
\newcommand{\bmat}[3]{\ensuremath{
																								\bigl \langle     #1   \bigr|    \, #2\,   \bigl|     #3   \bigr \rangle
                        									}
                     }
\newcommand{\submatB}[5]{\ensuremath{     {\vphantom{\ket{#1}}}_{#4} \mspace{- #5 mu} 
																																														 \mat{#1}{#2}{#3}
																																								}
                           } 
\newcommand{\bsubmatB}[5]{\ensuremath{     {\vphantom{\bket{#1}}}_{#4} \mspace{- #5 mu} 
																																														 \bmat{#1}{#2}{#3}
																																								}
                           }
\newcommand{\ket}[1]{\ensuremath{		\left| #1 \right> 
																																			  }
																									}
\newcommand{\bket}[1]{\ensuremath{		\bigl| #1 \bigr> 
																																			  }
																									}

\newcommand{\coh}[3]{\ensuremath{		\left( #1, #2 \right)_{#3} 
																																			  }
																									}

\newcommand{\overlap}[2]{\ensuremath{ 
																								\left \langle    #1 \vphantom{#2 } \,
                        \right| \left.   #2 \vphantom{#1}
                        \right \rangle
                        									}
                     }
\newcommand{\boverlap}[2]{\ensuremath{ 
																								\bigl \langle #1 \, \bigr| \bigl. #2 \bigr \rangle
                        									}
                     }
      
\newcommand{\bsuboverlapB}[4]{\ensuremath{     {\vphantom{\bket{#1}}}_{#3} \mspace{- #4 mu} 
																																														 \boverlap{#1}{#2} 
																																								}
                           }

\newcommand{\xOp}{ \ensuremath{  \hat{x}  }}
\newcommand{\pOp}{ \ensuremath{  \hat{p}  }}

\DeclareMathOperator{\Tr}{Tr}
\DeclareMathOperator{\sech}{sech}
\DeclareMathOperator{\cosech}{cosech}
\begin{document}
\begin{titlepage}
\begin{center}
\bfseries
GENERALISED HUSIMI FUNCTIONS:  ANALYTICITY AND INFORMATION CONTENT
\end{center}
\vspace{1 cm}
\begin{center}
D M APPLEBY 
\end{center}
\begin{center}
Department of Physics, Queen Mary and
		Westfield College,  Mile End Rd, London E1 4NS, UK
 \end{center}
\vspace{0.15 cm}
\begin{center}
  (E-mail:  D.M.Appleby@qmw.ac.uk)
\end{center}
\vspace{1.35 cm}
\vspace{0.6 cm}
\begin{center}
\textbf{Abstract}\\
\vspace{0.35 cm}
\parbox{10.5 cm }{  The analytic properties of a class of 
                    Generalised Husimi Functions are discussed, with particular
                    reference to the problem of state reconstruction.  The class consists of
                    the subset of W\'{o}dkiewicz's operational probability distributions for 
                    which the filter reference state is a squeezed vacuum state.
                    The fact that the function is analytic means that perfectly precise knowledge of
                    its values over any small region of phase space provides enough information to
                    reconstruct the density matrix. If, however, one only has imprecise knowledge of
                    its values, then the amplification of statistical errors which  occurs when one
                    attempts to carry out the continuation seriously limits the amount of information
                    which can be extracted.  To take account of this fact a distinction is made
                    between explicate, or experimentally accessible information, and information
                    which is only present in implicate, experimentally inaccessible form.  It is
                    shown that an explicate description of  various aspects of the system can
                    be found localised on  various 2 real dimensional surfaces in complexified
                    phase space.  In particular, the continuation of the function to the purely
                    imaginary part of complexified phase space provides an explicate description of
                    the Wigner function.                 }
\\     
\end{center}
\end{titlepage}
\section{Introduction}
\label{sec:  intro}
The purpose of this paper is to discuss the analytic properties of a class of
generalised Husimi functions, and the bearing that these properties have on the
problem of state reconstruction.

The functions we consider may be written
\begin{equation}
  Q_{\lambda \theta} (x, p)
= \frac{1}{\pi}
  \int dx' dp' \,
    \exp\left[- \frac{1}{\lambda^2} 
                 (x'_{\theta} - x^{\vphantom{t}}_{\theta})^2 
              - \lambda^2 (p'_{\theta} - p^{\vphantom{'}}_{\theta})^2
        \right] W(x',p')
\label{eq:  GenHusDef}
\end{equation}
where $W$ is the Wigner function, and where the notation $x_{\theta},p_{\theta}$
means
\begin{equation}
  \begin{pmatrix} x_{\theta} \\ p_{\theta} \end{pmatrix}
= \begin{pmatrix} \cos \theta & \sin \theta \\ - \sin \theta & \cos \theta
  \end{pmatrix}
  \begin{pmatrix} x \\ p \end{pmatrix}
\label{eq:  RotDef}
\end{equation}

The functions $Q_{\lambda \theta}$ belong to W\'{o}dkiewicz's class of 
operational probability 
distributions~\cite{Wod,Dav1}.  They are the distributions which result when
the filter reference state is chosen to be
an arbitrarily squeezed vacuum state~\cite{Squeeze}.
They have previously been discussed by Halliwell~\cite{Halli},
by W\"{u}nsche~\cite{Wuen}, and by W\"{u}nsche and Bu\v{z}ek~\cite{WuenBuz}.

In an earlier paper~\cite{self1} we discussed the physical interpretation of these
functions.  We showed that $Q_{\lambda \theta}$ describes the probability
distribution of measured values in the case when one makes a simultaneous,
retrodictively optimal measurement (using homodyne detection~\cite{Leon3,Leon}, for
example)   of
$\xOp_{\theta}$,
$\pOp_{\theta}$ to accuracies
$\pm
\frac{\lambda}{\sqrt{2}}$,
$\pm \frac{1}{\sqrt{2}\, \lambda}$ respectively [$\xOp_{\theta}$, $\pOp_{\theta}$
being defined in terms of $\xOp$, $\pOp$ by the obvious analogue of 
Eq.~(\ref{eq:  RotDef})].  This is a universal property:  it only
depends on the measurement being retrodictively optimal, and is otherwise
independent of the details of the particular measuring process employed.

It is natural to ask:  how much does one learn about the underlying quantum state
from a knowledge of $Q_{\lambda \theta}$?  Or, to put it another way:  what is the
information content of $Q_{\lambda \theta}$?  The answer to this question depends
on how accurately the function is known.  If it is known with \emph{perfect}
accuracy, then it provides all the information necessary to reconstruct the density
matrix~\cite{Dav1,WuenBuz,Mehta,Cahill,Prugo,Mizrahi,Dav3,Busch}.  However, the
process of reconstruction is extremely sensitive to small variations in $Q_{\lambda
\theta}$~\cite{leon2}.  Consequently, if one only has an \emph{approximate},
experimental  knowledge of the function, then it only provides a very limited
amount of information regarding the state.  One may express this by saying, that
although the transformation by which one obtains $Q_{\lambda \theta}$ from the
density matrix is invertible in point of mathematical principle, it is effectively
non-invertible in point of experimental practice.

These statements apply to $Q_{\lambda \theta}(x,p)$ regarded as a
function of two real variables.  However, $Q_{\lambda
\theta}(x,p)$ is in fact an analytic function of $x$, $p$ which continues to a
holomorphic function defined on the whole of complexified phase
space~\cite{Mehta,Cahill,Prugo,Dav3,Conn}.  Denote the continuation
$Q^{\mathrm{c}}_{\lambda
\theta}(x,p)$.  $Q^{\mathrm{c}}_{\lambda
\theta}(x,p)$ can itself be measured directly (by the method of quantum state 
sampling, for instance~\cite{Leon}). The question we
address in the following is:  how much information does one 
acquire by such measurements, either of the function as a whole,
or of its restriction to various surfaces in complexified phase space?

Of course, if one has perfectly accurate knowledge of $Q^{\mathrm{c}}_{\lambda
\theta}(x,p)$ on any small  region of complexified phase space having at least one
accumulation point then, because the function is holomorphic, that suffices to
determine the function everywhere, and it consequently suffices to determine the
density matrix.  However, from our point of view this fact is not very
interesting---for the reasons explained above.  We are primarily interested, not in
the information which can be extracted by some God-like being, endowed with
perfectly accurate knowledge regarding the values of the function in some region,
but rather with the information which is available to a real human experimenter,
whose knowledge is necessarily imperfect. We are accordingly led to make a
distinction between two different kinds of information, which we refer to as
``explicate'' and ``implicate'' (this terminology is  based on the
discussion in Bohm and Hiley~\cite{BohmBook}).  Explicate information is
information which is experimentally accessible.  Implicate information is
information which, though in some sense present, lies hidden in the fine
details of the analytic structure, and is therefore only accessible if one
has a perfectly accurate knowledge of the function.  In these terms the question
posed in the last paragraph may be rephrased:  what is the \emph{explicate} 
information
content of the function $Q^{\mathrm{c}}_{\lambda
\theta}(x,p)$?  And:  how is this information localised?

We begin in Section~\ref{sec:  invert} by deriving what we shall refer to as the
fundamental inversion formula.  In the case $\theta=0$
this formula was  obtained by Mehta and Sudarshan~\cite{Mehta}.  However, it
appears to us that it has not received the attention it deserves.  There do, of
course, exist a number of other formulae expressing the density matrix in terms of
the generalised Husimi 
function~\cite{Dav1,WuenBuz,Mizrahi,Dav3,Mehta2,Dav2,Reviews,Kano}.  However, the
formula derived in Section~\ref{sec:  invert} differs from all of these in that it
involves neither an integral nor an infinite sum.  The formula shows, that  
\emph{modulo} a Gaussian factor, the squeezed state density matrix elements (diagonal
\emph{and} off-diagonal) may actually be identified with the function
$Q^{\mathrm{c}}_{\lambda \theta}$. It follows, that having an approximate knowledge of
the values of
$Q^{\mathrm{c}}_{\lambda \theta}(x,p)$ for $x,p$ arbitrary complex, is equivalent
to knowing the density matrix, up to the same degree of approximation.  In other
words, $Q^{\mathrm{c}}_{\lambda \theta}$ (unlike $Q_{\lambda \theta}$) provides a 
fully explicate description of the density matrix.

In Section~\ref{sec:  SigAnal} we consider what happens if one attempts to use
the analyticity to infer the complete function $Q^{\mathrm{c}}_{\lambda \theta}$
starting from a knowledge of the restricted function $Q_{\lambda \theta}$.  We show that
the errors grow exponentially fast as one moves away from the real plane.  This means, 
that from an experimental point of view, 
the values which $Q^{\mathrm{c}}_{\lambda \theta}$ takes well away 
from the real plane are 
effectively independent of the values which it takes on the real plane.  In
Section~\ref{sec:  state} we go on to give a more precise definition of the concepts of
explicate and implicate information.

The explicate information content of $Q^{\mathrm{c}}_{\lambda \theta}$ is not
spread out evenly (so to speak).  Rather,
an explicate description of various different aspects of the system is to be found
concentrated on various  2 real dimensional subspaces of the 4 real
dimensional complexified phase space.  In Sections~\ref{sec:  xRep}--\ref{sec:  others}
we discuss this phenomenon.
In Section~\ref{sec:  xRep} we consider the problem of expressing the
$x_{\phi}$-representation density matrix elements in terms of $Q^{\mathrm{c}}_{\lambda
\theta}$.    In Section~\ref{sec:  change}
we consider the problem of expressing the function
$Q_{\lambda \theta}$ given by one choice of the parameters 
$\lambda$, $\theta$ in terms of the function 
$Q^{\mathrm{c}}_{\lambda' \theta'}$ given by some other choice. In Section~\ref{sec: 
others} we consider the problem of expressing other phase space distributions in terms
of $Q^{\mathrm{c}}_{\lambda
\theta}$.  In particular, we show that the continuation to the purely imaginary part 
of complexified phase space contains an explicate description of the Wigner function.  
We also
show that the existence and regularity of these distributions depends on the growth of
$Q^{\mathrm{c}}_{\lambda
\theta}$ as one moves away from the real plane.

Finally, in Section~\ref{sec:  comp} we show how the properties of $Q_{\lambda\theta}$
can be used to give an interesting illustration of the concept of complementarity.
\section{The Fundamental Inversion Formula}
\label{sec:  invert}
We begin by fixing some notation.  Let $\hat{a}_{\lambda \theta}$ be the annihilation
operator defined by 
\begin{equation*}
   \hat{a}_{\lambda \theta}^{\vphantom{\dagger}}  =
       \frac{1}{\sqrt{2}} 
       \left( \frac{1}{\lambda} \xOp_{\theta} 
             + i \lambda \pOp_{\theta}
        \right) 
\end{equation*}
where $\xOp_{\theta}$, $\pOp_{\theta}$ are the operators defined by
the obvious analogue of Eq.~(\ref{eq:  RotDef}).  Let $\ket{n}_{\lambda \theta}$ be a
normalised eigenstate of the number operator $\hat{a}_{\lambda \theta}^{\dagger}
\hat{a}_{\lambda
\theta}^{\vphantom{\dagger}}$ with eigenvalue $n$, and let 
$\ket{\coh{x}{p}{\lambda \theta}}$ be the squeezed state defined by
\begin{equation}
  \ket{  \coh{x}{p}{\lambda \theta} }
= \exp \left(- \tfrac{1}{2} \left| z\right|^2 \right)
  \sum_{n=0}^{\infty}
    \frac{z^{n}}{\sqrt{n!}}
    \ket{n}_{\lambda \theta}
\label{eq:  CohAsNoStes}
\end{equation}
where
   $z
=  \frac{1}{\sqrt{2}} \left( \frac{1}{\lambda} x_{\theta}
                       + i \lambda p_{\theta} \right)$.
Then the generalised Husimi function can be written
\begin{equation*}
  Q_{\lambda \theta} (x, p)
= \frac{1}{2 \pi}
   \bmat{\coh{x}{p}{\lambda \theta}
       }{\hat{\rho}
       }{\coh{x}{p}{\lambda \theta} }
\end{equation*}
where $\hat{\rho}$ is the density matrix.
More generally one may define, for any operator 
$\hat{A}$,
\begin{equation}
  A_{\lambda \theta} (x,p)
=     \bmat{\coh{x}{p}{\lambda \theta}
       }{\hat{A}
       }{\coh{x}{p}{\lambda \theta} }
\label{eq:  GenHusDefB}
\end{equation}
We will refer to this function as the generalised Husimi transform of
$\hat{A}$.  

The problem of reconstructing the density matrix from an exact
knowledge  of $Q_{\lambda \theta}$, and more generally the problem of
reconstructing an operator $\hat{A}$ from an exact knowledge of
$A_{\lambda\theta}$, comes down to the problem of using the diagonal matrix
elements in a squeezed state representation to reconstruct the
off-diagonal elements.  The possibility of such a reconstruction may,
at first sight, seem rather surprising.  It is certainly not
guaranteed just by the over-completeness of the set of squeezed states
(one can easily think of over-complete sets which do not have this
property).
What makes it possible is the analyticity of the transform.

There are, in fact, two different kinds of analyticity which are
relevant.  In the case of a pure state
Prugove\v{c}ki~\cite{Prugo} has used the analyticity of the 
Fock-Bargmann wave function~\cite{Bargmann,Perelomov} to show that the
Husimi function completely determines the state.  However, it is not
clear how to extend Prugove\v{c}ki's argument to deal with the case of
an arbitrary density matrix.  Also, Prugove\v{c}ki does not give a
formula for actually carrying out the inversion.

Rather more useful is the result obtained by Mehta and
Sudarshan~\cite{Mehta}.  Since Mehta and Sudarshan do not consider the
case $\theta\neq0$, and  since we feel that the result
deserves to be more widely known, we give the argument here.

In view of Eqs.~(\ref{eq:  CohAsNoStes}) and~(\ref{eq:  GenHusDefB})
we have
\begin{equation*}
    A_{\lambda \theta} (x,p)
=   \exp \left( - |z|^2 \right) 
    \sum_{n,m=0}^{\infty} \frac{ z^{* \, n}
z^{ m}}{\sqrt{ n! \, m!}} \ 
    \bsubmatB{n}{\hat{A}}{m}{\lambda \theta}{4}_{\lambda \theta}
\end{equation*}
Until now we have been assuming that $x$ and $p$ are both real. 
However, we can still make sense of this formula for complex $x$,
$p$.  Define
\begin{equation}
    z_{\pm}  =  
\frac{1}{\sqrt{2}} \left(
\frac{1}{\lambda} x_{\theta} \pm
i\lambda p_{\theta}\right)
\label{eq:  zpm}
\end{equation}
If $x$, $p$ are allowed to be complex then
$z_{\pm}$ are
independent complex variables.   Consider the expression
\begin{equation}
    A^{\mathrm{c}}_{\lambda\theta} (x,p)
=   \exp ( - z_{-} \,  z_{+} ) 
        \sum_{n,m=0}^{\infty} 
    \frac{ z_{-}^{ n} \, z_{ +}^{ m}}{\sqrt{ n! \, m!}} \ 
   \bsubmatB{n
  }{\hat{A}
  }{m
  }{\lambda\theta}{4}_{\lambda\theta}
\label{eq:  HTcontinue}
\end{equation}
It is straightforward to show~\cite{Cahill} that the series on the
right hand side of this equation converges if there exists a positive
constant $K$, and a constant $\alpha$ in the range $0\le\alpha<1$,
such that
\begin{equation*}
  \bigl\| \hat{A} 
         \ket{n}_{\lambda \theta} 
  \bigr\|
\le
\left(K n^{\alpha} \right)^n
\end{equation*}
for all $n$.  In that case Eq.~(\ref{eq:  HTcontinue}) gives the
continuation of $A_{\lambda \theta}$ to a holomorphic function defined
on the whole of $\mathbb{C}^2$.

Let us now see how we can use the continuation to invert the transform.
Consider the off-diagonal matrix element
$\bmat{\coh{x_1}{p_1}{\lambda \theta}
}{\hat{A}}{\coh{x_2}{p_2}{\lambda
\theta}}$. We have
\begin{align}
&   \bmat{\coh{x_1}{p_1}{\lambda\theta}
    }{\hat{A}
    }{\coh{x_2}{p_2}{\lambda\theta}}
\notag
\\
& \hspace{0.25 in}
=   \exp \left[
-\tfrac{1}{2}
 \left( |z_{1 }|^2 + |z_{2}|^2\right) \right]
    \sum_{n,m=0}^{\infty} 
    \frac{ z_{1 }^{* \, n} \, z_{2 }^{\phantom{*} \,m}}{\sqrt{n! \,
m!}} \ 
        \bsubmatB{n}{\hat{A}}{m}{\lambda}{4}_{\lambda}
\label{eq:  OffDiag}
\end{align}
where
\begin{equation*}
    z_{r}^{\vphantom{*}} 
 = \frac{1}{\sqrt{2}} 
\left( \frac{1}{\lambda} x_{r \theta} +
i\lambda p_{r \theta}\right)
\hspace{1.0 in}  r  = 1, 2
\end{equation*}
The points $(x_1,p_1)$ and $(x_2,p_2)$ both belong to $\mathbb{R}^2$.  
We now 
use them to define a single point 
$(x,p)$ in complexified phase space
\begin{equation*}
\begin{split}
    x & = \frac{1}{2} (x_1 + x_2) 
    + \frac{i}{2} 
    \sinh 2 \eta \sin 2 \theta \,
    (x_1 - x_2)
   - \frac{i}{2}
    \left(\cosh 2 \eta
     + \sinh 2 \eta \cos 2 \theta
    \right) (p_1 - p_2)
\\
    p & = 
    \frac{1}{2} (p_1 + p_2) 
    + \frac{i}{2 }
    \left( \cosh 2 \eta 
           - \sinh 2 \eta \cos 2 \theta
     \right)(x_1- x_2)
     -\frac{i}{2}
     \sinh 2 \eta \sin 2 \theta \,
     (p_1-p_2)
\end{split}
\end{equation*}
where we have set $\eta = \ln \lambda$.
We have
\begin{equation*}
\begin{aligned}
 x_{\theta} & = \cos \theta\, x + \sin
\theta \, p && = \frac{1}{2}
  \left(x_{1\theta}+x_{2\theta}
  \right)
  -\frac{i\lambda^2}{2}
  \left( p_{1\theta} - p_{2\theta}\right)
\\
p_{\theta} & = - \sin \theta \, x +
               \cos \theta \, p
&& = \frac{1}{2} 
     \left(p_{1 \theta}+ p_{2 \theta} 
     \right)
     + \frac{i}{2 \lambda^2}
     \left(x_{1\theta} - x_{2\theta}
     \right)
\end{aligned}
\end{equation*}
Consequently
\begin{equation*}
    z_{ +}^{\vphantom{*}} =
z_{2}^{\vphantom{*}}
\hspace{0.5 in} \text{and} \hspace{0.5 in} 
    z_{  -}^{\vphantom{*}} =
z_{1 }^{*}
\end{equation*}
where $z_{\pm}$ are the quantities defined by
Eq.~(\ref{eq:  zpm}).
Inserting these relations in 
Eq.~(\ref{eq:  OffDiag}) and comparing with
Eq.~(\ref{eq:  HTcontinue}) we deduce
\begin{equation*}
    \bmat{\coh{x_1}{p_1}{\lambda\theta}
    }{\hat{A}
    }{\coh{x_2}{p_2}{\lambda\theta}
    }
=   \exp 
   \left[ -\tfrac{1}{2}
   \left( |z_{1}^{\vphantom{*}}|^2 
         + |z_{2 }^{\vphantom{*}}|^2 
   \right) 
    +  z_{1 }^{*} \, z_{2 }^{\vphantom{*}}
        \right] \,
    A^{\mathrm{c}}_{\lambda\theta} (x,p)
\end{equation*}
which can alternatively be written
\begin{equation}
   A^{\mathrm{c}}_{\lambda\theta} (x,p)
=
    \frac{ 
    \bmat{\coh{x_1}{p_1}{\lambda \theta}
         }{\hat{A}
         }{\coh{x_2}{p_2}{\lambda\theta}
         } 
   }{
   \boverlap{\coh{x_1}{p_1}{\lambda\theta}
           }{\coh{x_2}{p_2}{\lambda\theta}
           }
    }
\label{eq:  QPrimeInv}
\end{equation}
Eq.~(\ref{eq:  QPrimeInv}) relates the off-diagonal matrix elements
to the analytic continuation of the diagonal elements.  We will refer
to it as the fundamental inversion formula.

If
$\theta =0$ and
$\lambda = 1$, and if
$\hat{A}$ is a density matrix, then the
expression on the left hand side is the
$Q$-function,
continued to complex values of $x$ and
$p$ (and \emph{modulo} a factor 
$\frac{1}{2 \pi}$);
while the expression on the right hand
side is the
$R$-representation~\cite{Glauber}
[\emph{modulo} a factor 
$\exp(- z_{1}^{*} 
\, z_{2}^{\vphantom{*}})$].
We see, therefore, that the
$R$-representation is essentially the same
thing as the analytic continuation of the
$Q$-function---as was originally noted
by Mehta and Sudarshan~\cite{Mehta}.

There do, of course, exist other ways 
of inverting the 
transform~\cite{Dav1,WuenBuz,Mizrahi,Dav3,Mehta2,Dav2,Reviews,Kano}.
However, these methods all involve, either
an integral, or else the calculation of an
infinite sum. The equation just derived is
more straightforward.  It shows, that
in order to invert the generalised Husimi
transform, all that one has to do is to
perform the analytic continuation, and
then to multiply by a Gaussian factor.

Eq.~(\ref{eq:  QPrimeInv}) also shows that the function
$Q^{\mathrm{c}}_{\lambda \theta}$ can be measured directly, by the 
method of quantum state sampling~\cite{Leon}.
\section{The Physical Significance of the Analyticity}
\label{sec:  SigAnal}
Eq.~(\ref{eq:  QPrimeInv}) shows that an exact specification of
$Q_{\lambda \theta}(x,p)$ for real $x$, $p$ suffices to
determine the density matrix.  Actually, a very much stronger statement
is true:  once
$Q^{\mathrm{c}}_{\lambda \theta}$ is exactly known on a region of
complexified phase space having at least one  accumulation point, then
that suffices to determine the function everywhere, and it consequently
suffices to determine the density matrix.  Alternatively, an exact
knowledge of the function together with all its derivatives at any
one point suffices to completely determine it at every other
point (as has been stressed by W\"{u}nsche and
Bu\v{z}ek~\cite{WuenBuz}).  The function might therefore be compared
with a hologram:  information sufficient to reconstruct the whole
picture is folded into every little piece of it.

$Q_{\lambda \theta}$ is a very intricately structured
object.   In particular, the analyticity 
means that
there exist subtle connections between the probability of 
finding the system
in one part  of phase space, and the probability of finding it another.
As an illustration of this point, specialise to the case $\theta=0$,
 and consider the following expression~\cite{Cahill} for
the quantity $\submatB{n}{\hat{\rho}}{n}{\lambda 0}{4}_{\lambda 0}$ 
 ($\ket{n}_{\lambda 0}$ being the number
state  defined at the beginning of the last section):
\begin{equation}
    \submatB{n}{\hat{\rho}}{n}{\lambda 0}{4}_{\lambda 0}
=   2 \pi \sum_{r=0}^{n} \frac{n!}{(n-r)! \ (r!)^2} \frac{1}{2^r} 
        \Biggl. \biggl[
       \Bigl( \lambda^2 \frac{\partial^2}{\partial x^2} 
            + \frac{1}{\lambda^2} \frac{\partial^2}{\partial p^2} 
       \Bigr)^r
       Q_{\lambda 0} (x,p)
       \biggr] \Biggr|_{x=p=0}
\label{eq:  NonLocal}
\end{equation}
$\submatB{n}{\hat{\rho}}{n}{\lambda 0}{4}_{\lambda 0}$ is the 
probability of finding the system in the state $\ket{n}_{\lambda 0}$. 
It  therefore tells us something about the probability of finding the
system in the vicinity of the surface $\frac{1}{\lambda^2} x^2 +
\lambda^2 p^2 = 2 n + 1$.  Yet in order  to calculate the probability we
have to evaluate $Q_{\lambda 0}$ and its derivatives, not in the
vicinity of this surface, but at the origin.  

The feature just evinced could be regarded as a kind of
non-locality.  It is somewhat reminiscent of the kind of non-locality
that is manifested by a violation of the Bell
inequalities~\cite{Bell}.

Bohm and Hiley~\cite{BohmBook} place much emphasis
on what, following Bohr~\cite{Bohr}, they describe as a property of
``wholeness'' possessed by quantum mechanical systems.  In their
discussion of this feature Bohm and Hiley also make use of a hologram
analogy.  They argue that the property of wholeness does not essentially
depend on assumptions peculiar to ``hidden-variables'' interpretations
of quantum mechanics, such as the Bohm interpretation.  On the contrary,
they maintain that it appears, in one guise or another, whatever the
interpretational scheme one adopts (it appears in the Copenhagen
interpretation favoured by Bohr, for example).  The analyticity of the
function
$Q^{\mathrm{c}}_{\lambda \theta}$ could be regarded as  another 
illustration of these ideas.

This phenomenon of ``wholeness'' is certainly  intriguing.
However, it is also elusive.  The non-local connections in the Bohm
interpretation cannot  be used for faster-than-light signalling. 
The practical significance of the kind of non-locality exemplified by
Eq.~(\ref{eq:  NonLocal}) is also very limited. It critically depends on
$Q^{\mathrm{c}}_{\lambda \theta}$ being known with perfect exactness.  
However,
$Q^{\mathrm{c}}_{\lambda
\theta}$ could in fact only ever be measured up to a certain non-zero
error,  which seriously restricts the  usefulness of the
analyticity.

To see this, consider the problem of calculating $Q^{\mathrm{c}}_{\lambda
\theta}$ at a point $(x+\eta, p+\zeta)$ using the expansion
\begin{equation*}
Q^{\mathrm{c}}_{\lambda \theta}(x+\eta, p+\zeta)
=\sum_{n,m=0}^{\infty} \frac{\eta^n \zeta^m}{n! m!}
\frac{\partial^{n+m}}{\partial x^n \partial p^m}
Q^{\mathrm{c}}_{\lambda
\theta} (x,p)
\end{equation*}
Suppose that the derivatives appearing in this sum have been measured
independently.  For the sake of example, suppose that the error in the
measurement of $\frac{\partial^{n+m}}{\partial x^n \partial p^m}
Q^{\mathrm{c}}_{\lambda
\theta}$ is $\pm \lambda^{m-n} \sigma$, where $\sigma$ is a
dimensionless constant.  Let $\Delta Q$ be the error in
$Q^{\mathrm{c}}_{\lambda \theta}(x+\eta,p+\zeta)$, as calculated using
the above formula.  Then
\begin{equation*}
  \Delta Q
= \sigma \left( \sum_{n,m=0}^{\infty} \frac{|\eta|^{2 n} |\zeta|^{2
m}}{(n! m!)^2}
\lambda^{2(m - n)}\right)^{\frac{1}{2}}
= \sigma \Bigl[ I_{0} \left(2 \lambda^{-1} |\eta|\right) I_{0} \left(2
\lambda |\zeta|\right)\Bigr]^{\frac{1}{2}}
\end{equation*}
where $I_{0}$ is a Bessel function of imaginary argument~\cite{Grad}.
Asymptotically~\cite{Grad}
\begin{equation*}
  \Delta Q \sim \frac{1}{2} \sigma \left(\pi^2 |\eta|
|\zeta|\right)^{-\frac{1}{4}} \exp\left(\lambda^{-1} |\eta| +\lambda
|\zeta|\right)
\end{equation*}
It can be seen that the error grows exponentially  as one moves away
from the point $(x,p)$.  If, instead of direct measurements of the
derivatives, one only had imprecise information regarding the values of
the function  itself in some small region surrounding the point $(x,p)$,
from which the values of the derivatives had to be estimated, then one
would expect the error to grow even more rapidly.

We see that the situation is not really so  different from the
case of a function which is not  analytic.  Given an
experimental determination of the function in some region, one can use
the analyticity to infer something about its behaviour a small distance
outside that region.  However, one cannot go very far before the data is
swamped by the errors.

Although they  are not independent in point of mathematical
principle, it may be said that the values which $Q^{\mathrm{c}}_{\lambda
\theta}$ takes in well-separated regions of complexified phase space are
effectively independent in point of experimental practice.

The situation here could be compared with the case of a classically
chaotic deterministic system.  In principle the state of such a system
at any one time suffices to completely determine its state at every other
time.  However, the practical usefulness of this fact is very limited. 
The information is in some sense present, but mostly not in a form which
it is possible to access.
\section{Explicate v. Implicate Information Content}
\label{sec:  state}
In view of the discussion in the last section we are led to distinguish
between two different kinds of information which, following Bohm and
Hiley~\cite{BohmBook}, we will refer to as ``explicate'' and
``implicate'' (although we do not use these words in quite the same sense
as Bohm and Hiley).  Explicate information is information which
approximately survives the process of making inexact measurements. 
Implicate information is information which is experimentally
inaccessible.

In order to make these notions a little more precise, consider a
physical quantity which is given by  an expression of the form
\begin{equation*}
   \int_{D} f(x,p) Q^{\mathrm{c}}_{\lambda \theta} (x, p) d \mu
\end{equation*}
In this expression the domain of integration $D$ might be a 4 real
dimensional volume in complexified phase space, or it might be a lower
dimensional surface (such as the real plane).  $f$ might be an ordinary
function, or it might only be defined in a distributional sense.

Suppose that $f$ is an ordinary function.  In that case the error in the
integral is of a comparable order to the error in
$Q^{\mathrm{c}}_{\lambda \theta}$.  The same is true for certain, not
too singular distributions.  It is true if $f$ is a $\delta$ function, for
example. In such a case we will say that $Q^{\mathrm{c}}_{\lambda
\theta}$, restricted to the region $D$, contains explicate information
regarding the physical quantity in question.

Suppose, on the other hand, that $f$ is a very singular
distribution.  Suppose, for example, that $f$ involves many high order
derivatives of the $\delta$ function.  In that case, although the
integral may still be of some theoretical interest, it cannot easily be
used to make practical calculations based on inexact experimental data
concerning the function  $Q^{\mathrm{c}}_{\lambda
\theta}$.  We will say that $Q^{\mathrm{c}}_{\lambda
\theta}$, restricted to the region $D$, only contains implicate
information about the physical quantity in question.

The distinction between ``explicate'' and ``implicate'' is qualitative,
and it should not be taken in too absolute a sense.  For instance,  one
cannot entirely rule out the possibility of measuring all the quantities
on the right hand side of Eq.~(\ref{eq:  NonLocal}), and using them to
make a good estimate of $\submatB{n}{\hat{\rho}}{n}{\lambda
0}{4}_{\lambda 0}$.  However, it would obviously be very difficult if
$n$ is large.

As an illustration of these concepts, consider the problem of
calculating the expectation value of an operator $\hat{A}$.  Using
Eq.~(\ref{eq:  QPrimeInv}) we have
\begin{align}
    \Tr ( \hat{A}\hat{\rho} )
& = \frac{1}{4 \pi^2} \int dx_1 dp_1 dx_2 dp_2 \,
       \bmat{\coh{x_1}{p_1}{\lambda \theta}
           }{\hat{A}}{\coh{x_2}{p_2}{\lambda \theta}} \,
       \bmat{\coh{x_2}{p_2}{\lambda \theta}
           }{\hat{\rho}}{\coh{x_1}{p_1}{\lambda \theta}}
\notag
\\
& = \frac{2}{\pi} \int d^2 x \, d^2 p \,
       \exp \left(- \frac{2}{\lambda^2} x_{\mathrm{I}\theta}^2 
                  - 2 \lambda^2 p_{\mathrm{I} \theta}^2
            \right)
       A^{\mathrm{c}}_{\lambda\theta} (x^*, p^*)
Q^{\mathrm{c}}_{\lambda\theta} (x,p)
\label{eq:  FullHusExpect}
\end{align}
In this expression $x_{\mathrm{I}}$, $p_{\mathrm{I}}$ denote the
imaginary parts of the complex variables
$x$, $p$.  $x_{\mathrm{I}\theta}$, $p_{\mathrm{I}\theta}$ are defined
in terms of $x_{\mathrm{I}}$, $p_{\mathrm{I}}$ as in 
Eq.~(\ref{eq:  RotDef}).  The integral is taken over the whole of
complexified phase space.  
It can be seen that
$Q^{\mathrm{c}}_{\lambda \theta}$ contains  explicate information about
$\langle \hat{A} \rangle$ whenever 
$A^{\mathrm{c}}_{\lambda \theta}$ is defined as an ordinary
function---which is the case for most operators of physical interest.

Suppose, on the other hand, that one wants to calculate 
$\langle\hat{A}\rangle$ using only the values of the
restricted function $Q_{\lambda \theta}$.
Inspection of Eq.~(\ref{eq:  FullHusExpect}) shows, that due to the
presence of the Gaussian factor, 
the integrand will be strongly peaked
about the real plane, provided that 
$A^{\mathrm{c}}_{\lambda\theta} (x^*, p^*)$ does not
grow too rapidly with increasing 
$x_{\mathrm{I}}^{\vphantom{n}}$, $p_{\mathrm{I}}^{\vphantom{n}}$.  
In that case we may 
expand  $A^{\mathrm{c}}_{\lambda \theta} (x^*, p^*)$, and
$Q^{\mathrm{c}}_{\lambda \theta} (x, p)$ 
about the point 
$(x_{\mathrm{R}}^{\vphantom{n}}, p_{\mathrm{R}}^{\vphantom{n}})$ (where
$x_{\mathrm{R}}^{\vphantom{n}}$, $p_{\mathrm{R}}^{\vphantom{n}}$ denote
the real parts of $x$, $p$): 
\begin{align*}
    \Tr (\hat{A} \hat{\rho} )
& = \frac{2}{\pi} \sum_{n,n',m,m'=0}^{\infty} 
    \int d^2 x \, d^2 p\, 
    \exp\left (-\frac{2}{\lambda^2} x_{\mathrm{I}\theta}^{2} 
               - 2 \lambda^2 p_{\mathrm{I}\theta}^{2}
\right) 
    \frac{ (-1)^{n+m} \, i^{n+n'+m+m'}}{n!\,n'!\,m!\,m'!}
\notag
\\
& \hspace{0.75 in} \times
    x_{\mathrm{I}}^{n+n'} p_{\mathrm{I}}^{m+m'}
    \frac{\partial^{n+m} }{\partial x_{\mathrm{R}}^n \, \partial p_{\mathrm{R}}^m}
    A_{\lambda\theta} (x_{\mathrm{R}}^{\vphantom{n}},
p_{\mathrm{R}}^{\vphantom{n}}) \,
    \frac{\partial^{n'+m'} }{\partial x_{\mathrm{R}}^{n'} \, \partial p_{\mathrm{R}}^{m'}} 
    Q_{\lambda \theta} (x_{\mathrm{R}}^{\vphantom{n}}, p_{\mathrm{R}}^{\vphantom{n}})
\end{align*}
One finds
\begin{equation}
\Tr (\hat{A} \hat{\rho} )
= \int dx_{\mathrm{R}} dp_{\mathrm{R}}\,
          A^{\mathrm{a}}_{\lambda \theta} (x_{\mathrm{R}},
p_{\mathrm{R}})
           Q_{\lambda \theta} (x_{\mathrm{R}}, p_{\mathrm{R}})
\label{eq:  ExpInGAH}
\end{equation}
where $A^{\mathrm{a}}_{\lambda \theta}$ is the generalised anti-Husimi 
transform~\cite{Dav2,Reviews},  defined by
\begin{equation}
      A^{\mathrm{a}}_{\lambda \theta} (x_{\mathrm{R}}^{\vphantom{n}},
p_{\mathrm{R}}^{\vphantom{n}} ) =    \exp \left[ - \frac{1}{2} 
                   \left( \lambda^2\frac{\partial^2}{\partial x_{\mathrm{R}\theta}^2}
                         + \frac{1}{\lambda^2 }\frac{\partial^2}{\partial p_{\mathrm{R}\theta}^2}
                   \right)
          \right]
     A_{\lambda \theta} (x_{\mathrm{R}}^{\vphantom{n}},
p_{\mathrm{R}}^{\vphantom{n}} )
\label{eq:  GAHDef}
\end{equation}
There are, of course, easier ways~\cite{Dav1} of deriving 
Eq.~(\ref{eq:  ExpInGAH}).  The advantage of the derivation just
indicated is that it brings out the connection between  
Eqs.~(\ref{eq:  FullHusExpect}) and~(\ref{eq:  ExpInGAH}).  It also
brings out the fact that the existence of
$A^{\mathrm{a}}_{\lambda \theta}$ as an ordinary function depends on the
growth of
$A^{\mathrm{c}}_{\lambda \theta}$ as one moves away from the real plane. 
We will return to this point in Section~\ref{sec:  others}.

If $A^{\mathrm{a}}_{\lambda \theta}$ exists as an ordinary function, or
as a not too singular distribution, then $Q_{\lambda \theta}$ provides
us with explicate information regarding $\langle\hat{A}\rangle$. 
However, $A^{\mathrm{a}}_{\lambda \theta}$ is in fact often very
singular.  The conclusion is, therefore, that whereas
$Q^{\mathrm{c}}_{\lambda \theta}$ provides us with an explicate
description of the quantum state, the description provided by 
$Q_{\lambda \theta}$ is, for the most part, only implicate.

$Q_{\lambda\theta}$ is the probability distribution describing the
outcome of a retrodictively optimal determination of phase space
location~\cite{self1}.  It might therefore be compared with the 
function $|\overlap{x}{\psi}|^2$.  The values which
$Q^{\mathrm{c}}_{\lambda \theta}$ takes away from the real plane provide
us with the additional, explicate information needed to calculate the
expectation value of an arbitrary observable.  They might therefore be
compared with the phase of the function $\overlap{x}{\psi}$.
\section{Connection with the $x_{\phi}$-Representation Matrix Elements}
\label{sec:  xRep}
The advantage of Eq.~(\ref{eq:  ExpInGAH}) is that it only requires us
to integrate over two real variables, unlike Eq.~(\ref{eq: 
FullHusExpect}) which requires us to integrate over four.  The
disadvantage is that $A^{\mathrm{a}}_{\lambda\theta}$ is
often extremely singular.  It is natural to ask whether one might do
better by integrating over some other surface, instead of the
real plane.  In the next three sections we investigate this question. 
We begin by identifying a surface, having two real dimensions, which
contains an explicate description of the matrix elements
\begin{equation*}
  \bsubmatB{x_1}{\hat{\rho}}{x_2}{\phi}{4}_{\phi}
\end{equation*}
(where $\ket{x}_{\phi}$ is an eigenket of 
$\hat{x}_{\phi}$ with eigenvalue
$x$).  The result we derive is a generalisation of the one proved by 
Davidovi\'{c} and Lalovi\'{c}~\cite{Dav3}.

It is readily shown~\cite{Squeeze,Leon}
\begin{align*}
    \bsuboverlapB{x'}{\coh{x}{p}{\lambda \theta}}{\phi}{4}
& = \boverlap{x'}{\coh{x_{\phi}}{p_{\phi}}{\lambda \delta}}
\\
& = \left( \frac{e^{- i \delta}}{ \sqrt{\pi} \left( \lambda \cos \delta - i \lambda^{-1} \sin \delta\right)}
     \right)^{\frac{1}{2}}
\\
& \hspace{0.25 in} \times
     \exp \left[ - \frac{  \lambda^{-1} \cos \delta - i \lambda \sin \delta
                       }{  2\left(\lambda \cos\delta - i \lambda^{-1} \sin \delta\right)
                        }
                   (x' - x_{\phi})^2 + i p_{\phi} x' - \frac{1}{2} p_{\phi} x_{\phi}
           \right]
\end{align*}
where $\delta = \theta - \phi$.
Consequently
\begin{align}
\notag
\\
 Q_{\lambda \theta} (x,p)
& = \frac{1}{\pi^{\frac{3}{2}} \sqrt{ b}}
    \int dx' dy' \,
       \exp\left[- \frac{1}{b} 
                 \left( (x'-x_{\phi})^2 + y'\vphantom{y}^{2}\right)
            \right.
\notag
\\
& \hspace{1.5 in}     
           \left.
                 - \frac{ 2 i c}{b} y' (x'- x_{\phi})
                 + 2 i p_{\phi} y'
           \right]
     \bsubmatB{x'-y'}{\hat{\rho}}{x'+y'}{\phi}{4}_{\phi}
\label{eq:  QInXthRepB}
\end{align}
where
\begin{equation*}
\begin{split}
  b & = \lambda^2 \cos^2 \delta + \lambda^{-2} \sin^2 \delta \\
  c & = - \left( \lambda^2 - \lambda^{-2}\right) \sin \delta \cos \delta
\end{split}
\end{equation*}
Continuing to complex values of $x$ and $p$ we find
\begin{align}
& \exp(-b u^2) Q^{\mathrm{c}}_{\lambda \theta} \,
    \left( - v \sin \phi  + i u d, v \cos \phi  + i u f \right)
\notag
\\
& \hspace{0.2 in} 
=
  \frac{1}{\pi^{\frac{3}{2}} \sqrt{b}}
  \int dx dy \,
      \exp \left[ 2 i (u x + v y)\right]
\notag
\\
& \hspace{1.5 in} \times
      \exp \left[ - \tfrac{1}{b} (x^2 + 2 i c x y + y^2)\right]
     \bsubmatB{x-y}{\hat{\rho}}{x+y}{\phi}{4}_{\phi}
\label{eq:  QInXthRep}
\end{align}
where
\begin{equation*}
\begin{aligned}
  d & = b \cos \phi + c \sin \phi & 
      & = \lambda^2 \cos \theta \cos \delta + \lambda^{-2} \sin \theta \sin\delta \\
  f & = b \sin \phi - c \cos \phi & 
     & = \lambda^2 \sin \theta \cos \delta - \lambda^{-2} \cos \theta \sin \delta
\end{aligned}
\end{equation*}
Inverting the Fourier transform gives
\begin{align}
&        \exp \left[ - \tfrac{1}{b} (x\vphantom{x}^2 + 2 i c x y + y \vphantom{y}^2)\right]
     \bsubmatB{x-y}{\hat{\rho}}{x +y}{\phi}{4}_{\phi}
\notag
\\
& \hspace{0.2 in}
=  \left(\frac{b}{\pi}\right)^{\frac{1}{2}}
   \int du dv \,
      \exp \left[ - 2 i (u x + vy)\right]
\notag
\\
& \hspace{1.5 in} \times
      \exp(-b u^2) Q^{\mathrm{c}}_{\lambda \theta} \,
    \left( - v \sin \phi  + i u d, v \cos \phi  + i u f \right)
\label{eq:  xthRepInv}
\end{align}
which is the desired result.

By essentially the same argument one can derive an analogous formula
relating the $x_{\phi}$-representation matrix elements of an arbitrary
operator
$\hat{A}$ to the function $A^{\mathrm{c}}_{\lambda \theta}$.
\section{$Q_{\lambda' \theta'}$ in terms of $Q^{\mathrm{c}}$}
\label{sec:  change}
Using similar means one may obtain an
explicate description of the function $Q_{\lambda \theta}$ by
integrating $Q^{\mathrm{c}}_{\lambda' \theta'}$ over a 
surface having 2 real dimensions.
However, the general formula is rather complicated.  We will
therefore confine ourselves to showing how 
$Q_{\lambda \theta}$ may be expressed in terms of the analytic
continuation of the $Q$-function, $Q^{\mathrm{c}}=Q^{\mathrm{c}}_{10}$.

Suppose that $\lambda <1$.  Eq.~(\ref{eq:  QInXthRep}) gives
\begin{align*}
&  \exp(-\lambda^{-2} u^2)
   Q^{\mathrm{c}}_{\lambda \theta} (-v \sin \theta + i u \cos \theta, v \cos
\theta + i u \sin \theta)
\\
& \hspace{0.2 in}
=  \frac{1}{\pi^{\frac{3}{2}} \lambda}
  \int dx dy \, \exp \bigl[ 2 i (\lambda^{-2} u x + v y)\bigr] \,
  \exp \bigl[ - \lambda^{-2} (x^2 + y^2)\bigr] \,
  \bsubmatB{x-y}{\hat{\rho}}{x+y}{\theta}{3}_{\theta}
\end{align*}
while  Eq.~(\ref{eq:  xthRepInv}) gives
\begin{align*}
& \exp \left[ - (x^2+y^2)\right] \,
  \bsubmatB{x-y}{\hat{\rho}}{x+y}{\theta}{3}_{\theta}
\\
& \hspace{0.2 in}
= \frac{1}{\sqrt{\pi}}
  \int du' dv' \,
      \exp \left[ - 2 i (u' x + v' y)\right] \,
      \exp (- u^2)
\\
& \hspace{2.0 in} \times
      Q^{\mathrm{c}} (-v' \sin \theta + i u' \cos \theta, v' \cos \theta + i u'
\sin \theta )
\end{align*}
Combining these formulae we obtain

\begin{align}
&   \exp{\left[-\frac{1}{1-\lambda^2} x_{\theta}^2\right]}
Q_{\lambda\theta}(x,p) 
\notag
\\
& \hspace{0.25 in}
  = 
  \frac{\lambda}{\pi(1-\lambda^2)}
  \int du' dv' \,
     \exp{\left[-\frac{2 i}{1-\lambda^2} u' x_{\theta}
               -\frac{1}{\lambda^{-2} -1} (v'-p_{\theta})^2\right]}
\notag
\\
& \hspace{1.0 in}
  \times
  \exp{\left[-\frac{1}{1-\lambda^2} {u'}^{2}\right]}
   Q^{\mathrm{c}} (-v' \sin \theta + i u' \cos \theta,
                    v' \cos \theta+ i u'\sin \theta)
\label{eq:  GenHusInQLess}
\end{align}
If $\lambda > 1$ we have
\begin{align}
&   \exp{\left[-\frac{1}{1-\lambda^{-2}} p_{\theta}^2\right]}
Q_{\lambda\theta}(x,p) 
\notag
\\
& \hspace{0.25 in}
  = 
  \frac{\lambda^{-1}}{\pi(1-\lambda^{-2})}
  \int du' dv' \,
     \exp{\left[\frac{2 i}{1-\lambda^{-2}} u' p_{\theta}
               -\frac{1}{\lambda^{2} -1} (v'-x_{\theta})^2\right]}
\notag
\\
& \hspace{1.0 in}
  \times
  \exp{\left[-\frac{1}{1-\lambda^{-2}} {u'}^{2}\right]}
   Q^{\mathrm{c}} (v' \cos \theta + i u' \sin \theta,
                    v' \sin \theta - i u'\cos \theta)
\label{eq:  GenHusInQGreat}
\end{align}

$Q$ describes the result~\cite{self1} of measuring $x_{\theta}$, $p_{\theta}$ both
to a retrodictive  accuracy of $\pm\frac{1}{\sqrt{2}}$.  Eq.~(\ref{eq: 
GenHusInQLess}) shows that continuing to one part of complexified phase space
gives a more accurate explicate description of $x_{\theta}$. 
Eq.~(\ref{eq:  GenHusInQGreat}) shows that  continuing to a different part of
complexified phase space gives a more accurate explicate description of
$p_{\theta}$.  An explicate description of the result of every possible measurement
is to be found somewhere in complexified phase space.

Analogous formulae hold for the generalised Husimi transform of an arbitrary
operator $\hat{A}$.
\section{The Generalised $s$-Transform}
\label{sec:  others}
The generalised $s$-transform of an operator $\hat{A}$ is defined by 
\begin{equation*}
  A^{(s)}_{\lambda \theta} (x, p)
= \frac{1}{2 \pi} \int dx' dp' \,
     \exp \left[ i (px'-xp') 
                 + \frac{s}{4} (\lambda^2 p'\vphantom{p}_{\theta}^{2}
                                + \lambda^{-2} x'\vphantom{x}_{\theta}^{2})
          \right]
     \Tr \left( \hat{D}_{x'p'} \hat{A} \right)
\end{equation*}
where $\hat{D}_{xp}=\exp[i(p\xOp - x \pOp)]$ is the displacement
operator.

If $s>-1$ it is straightforward to show
\begin{align}
  \hat{A}_{\lambda \theta} (x, p)
& =  \frac{1}{\pi (1+s)}
   \int dx' dp' \,
      \exp \left[ - \frac{1}{(1+s)} \left( \lambda^{-2} (x'_{\theta} - x_{\theta})^2
                  + \lambda^2 (p'_{\theta}-p_{\theta})^2 \right)
           \right]
\notag
\\
& \hspace{2.5 in} \times
      A^{(s)}_{\lambda\theta} (x',p')
\label{eq:  GensTransToH}
\end{align}
This formula can be inverted using
\begin{equation}
   A^{(s)}_{\lambda \theta} (x,p)
=  \exp \left[ - \frac{1+s}{4} \left( \lambda^2 \frac{\partial^2}{\partial x_{\theta}^{2}}
                                      + \frac{1}{\lambda^2} \frac{\partial^2}{\partial p_{\theta}^{2}}
                               \right)
        \right]
   A_{\lambda \theta} (x, p)
\label{eq:  HTransToGens}
\end{equation}
It is to be observed, however, that the expression on the right hand side
of Eq.~(\ref{eq:  HTransToGens}) contains
derivatives of all orders, which means  it is
exploiting information that is only present in implicate  form.
If one wants a formula which only depends on information present in explicate
form it is necessary to  make use of the analytic continuation.

Continuing to imaginary values of $x$ and $p$ 
in Eq.~(\ref{eq:  GensTransToH}) gives,
after rearranging, 
\begin{align*}
& \exp\left[ - \frac{1}{1+s} \left( \lambda^{-2} x^2 + \lambda^2 p^2 \right)
      \right]
  A^{\mathrm{c}}_{\lambda \theta} (i x_{-\theta}, i p_{-\theta})
\\
& \hspace{0.25 in}
=  \frac{1}{\pi (1+s)}
   \int dx' dp' \,
       \exp \left[ \frac{2 i}{(1+s)} (\lambda^{-2} x x' + \lambda^{2} p p')
            \right] 
\\
& \hspace{1.5 in} \times
       \exp \left[ - \frac{1}{1+s} \left( \lambda^{-2} x'\vphantom{x}^2 + \lambda^2 p'\vphantom{p}^2 \right)
            \right]
       A^{(s)}_{\lambda \theta} (x'_{-\theta},p'_{-\theta})
\end{align*}
Inverting the Fourier transform and changing the variables of integration gives
\begin{align}
&   \exp \left[ - \frac{1}{1+s} \left( \lambda^{-2} x_{\theta}^2 + \lambda^2 p_{\theta}^2 \right)
            \right]
       A^{(s)}_{\lambda \theta} (x,p)
\notag
\\
& \hspace{0.25 in}
=  \frac{1}{\pi (1+s)}
   \int dx' dp' \,
       \exp \left[ - \frac{2 i}{(1+s)} (\lambda^{-2} x_{\theta} x'_{\theta} 
                                      + \lambda^{2} p_{\theta} p'_{\theta})
            \right] 
\notag
\\
& \hspace{1.5 in} \times
       \exp \left[ - \frac{1}{1+s} \left( \lambda^{-2} x'\vphantom{x}^{2}_{\theta} 
                                        + \lambda^2 p'\vphantom{p}^{2}_{\theta} \right)
            \right]
       A^{\mathrm{c}}_{\lambda \theta} (ix',ip')
\label{eq:  HTransToGensB}
\end{align}
If $s=0$ and $\hat{A}$ is the density matrix this becomes
\begin{align}
&   \exp \left[ - \left( \lambda^{-2} x_{\theta}^2 + \lambda^2 p_{\theta}^2 \right)
            \right]
       W (x,p)
\notag
\\
& \hspace{0.25 in}
=  \frac{1}{\pi }
   \int dx' dp' \,
       \exp \left[ - 2 i (\lambda^{-2} x_{\theta} x'_{\theta} 
                                      + \lambda^{2} p_{\theta} p'_{\theta})
            \right] 
\notag
\\
& \hspace{1.5 in} \times
       \exp \left[ - \left( \lambda^{-2} x'_{\theta}\vphantom{x}^{2} 
                                        + \lambda^2 p'_{\theta}\vphantom{p}^{2} \right)
            \right]
       Q^{\mathrm{c}}_{\lambda \theta} (ix',ip')
\label{eq:  WigInGenQ}
\end{align}
We see that the continuation of $Q_{\lambda \theta}$ to the purely imaginary part of 
complexified phase space contains an explicate description of the Wigner function.

Eq.~(\ref{eq:  HTransToGensB}) also implies a convenient criterion for the 
existence and regularity properties of the generalised $s$-transform, in terms of
the growth  of $A^{\mathrm{c}}_{\lambda \theta}$ as one moves away from the real
plane. Specifically:  $A^{(s)}_{\lambda \theta} (x, p)$ 
exists as a tempered distribution if and only if
$\exp\left[-\frac{1}{1+s} (\lambda^{-2} x_{\theta}^{2} +\lambda^{2} p_{\theta}^{2})\right]
A^{\mathrm{c}}_{\lambda \theta}(ix,ip)$ exists as a tempered distribution
($x$ and $p$ both real).

In view of Eq.~(\ref{eq:  QPrimeInv}) the condition may be rephrased:  the necessary 
and sufficient condition for $A^{(s)}_{\lambda \theta} (x, p)$ to exist 
as a tempered distribution is that
\begin{equation*}
\exp\left[ \frac{s}{1+s} (\lambda^{-2} x_{\theta}^{2} + \lambda^{2} p_{\theta}^{2} )\right]
\bmat{\coh{-x}{-p}{\lambda \theta}}{\hat{A}}{\coh{x}{p}{\lambda \theta}}
\end{equation*}
exists as a tempered distribution.  In particular, the
generalised $s$-transform  exists as a tempered distribution whenever 
the operator $\hat{A}$ is local,
in the sense that it takes localised wave packets into wave packets which are again
localised, in the same region of phase space.
\section{Complementarity}
\label{sec:  comp}
We conclude by showing how these considerations can be used to 
illustrate the concept of complementarity.

Suppose one took a \emph{classical} system characterised by the classical
distribution $\Gamma$ and measured the observables
$x_{\theta}$, $p_{\theta}$ to accuracies $\pm \frac{\lambda}{\sqrt{2}}$,
$\pm\frac{1}{\sqrt{2}\lambda}$ respectively.  The probability distribution
for the values of $x$, $p$ as calculated  from the measured values of
$x_{\theta}$, $p_{\theta}$ would be
\begin{equation}
   \rho_{\lambda \theta} (x, p)
=  \frac{1}{\pi} \int dx' dp' \,
      \exp \left[ - \lambda^{-2} (x'\vphantom{x}_{\theta} - x_{\theta})^2
                  - \lambda^2 (p'\vphantom{p}_{\theta} - p_{\theta})^2
           \right]
      \Gamma (x', p')
\label{eq:  ClassDist}
\end{equation}
Comparing the defining equation for $Q_{\lambda\theta}$, Eq.~(\ref{eq: 
GenHusDef}), with this equation it can be seen that the only difference consists
in the replacement of the classical distribution $\Gamma$ with the Wigner function
$W$.  It is natural to ask:  can one infer that the underlying
distribution is $W$  rather than $\Gamma$ from experimental determinations of the
functions
$Q_{\lambda \theta}$?
The answer to this question is that the non-classical nature of
the  $Q_{\lambda\theta}$ does not become experimentally manifest for any single
pair of values of 
$\lambda$ and
$\theta$.  However, it does show up if one makes measurements for many different
values of these parameters.

To see this, consider Eq.~(\ref{eq:  QInXthRepB}) for the case $\phi = \theta$:
\begin{align*}
  Q_{\lambda \theta}(x,p)
& = \frac{1}{\pi^{\frac{3}{2}} \lambda}
    \int dx' dp' \,
        \exp \left[ - \frac{1}{\lambda^2} \left( (x' - x_{\theta})^2 + y'\vphantom{y}^2 \right)
                    + 2 i p_{\theta} y'
             \right]
\\
& \hspace{2.0 in} \times
        \submatB{x'-y'}{\hat{\rho}}{x'+y'}{\theta}{3}_{\theta}
\end{align*}
In the limit as $\lambda \rightarrow 0$
\begin{align*}
   \frac{1}{\sqrt{\pi} \lambda } \, \exp \left[ - \frac{1}{\lambda^2} (x' - x_{\theta})^2 \right]
\approx \delta (x' - x_{\theta})
\\
\intertext{and}
  \frac{1}{\sqrt{\pi} \lambda } \, 
  \exp \left[ - \frac{1}{\lambda^2} y'\vphantom{2} + 2 i p_{\theta} y' \right]
\approx
  \exp\left[ - \lambda^2 p_{\theta}^2\right] \delta (y')
\end{align*}
Consequently
\begin{equation}
  Q_{\lambda \theta}(x,p)
\approx \frac{\lambda}{\sqrt{\pi}} \exp \left[ -\lambda^2 p_{\theta}^2 \right]
       \submatB{x_{\theta}}{\hat{\rho}}{x_{\theta}}{\theta}{3}_{\theta}
\label{eq:  GenHusAsymp}
\end{equation}
in the small $\lambda$ limit.  An experimental determination of the  functions
$\submatB{x_{\theta}}{\hat{\rho}}{x_{\theta}}{\theta}{3}_{\theta}$ provides
one with enough information for an approximate reconstruction of the 
quantum state~\cite{Leon3,Leon,Vogel,TomExp,Symplect}. 
It follows, that an experimental determination of 
the  $Q_{\lambda \theta}(x,p)$ (with
$x$, $p$ both real) also provides one with enough
information for such a reconstruction, provided that the functions are known
for sufficiently small $\lambda$, and sufficiently 
many different values of $\theta$.

\begin{figure}
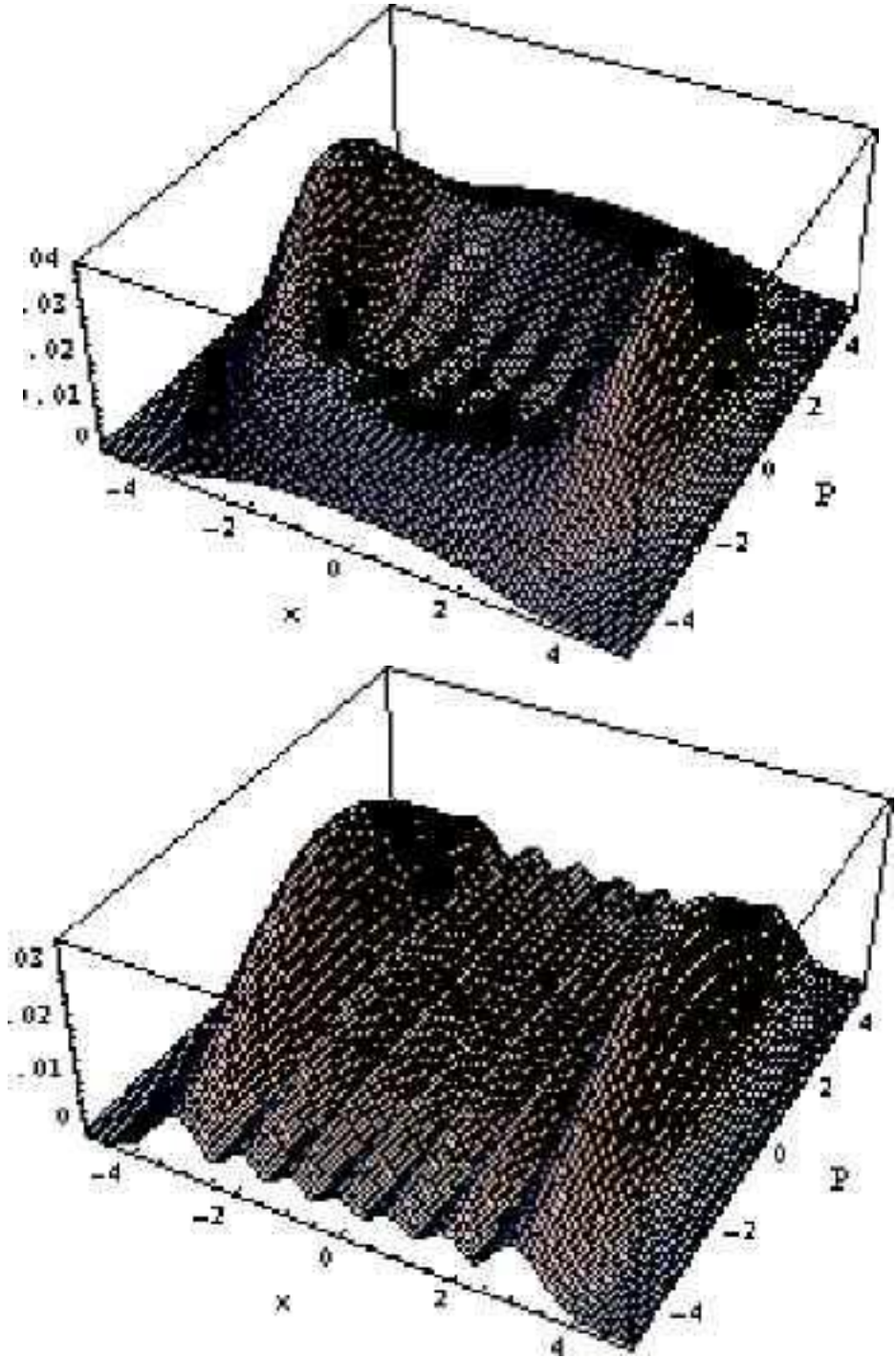

\includegraphics{HusFig3abm.EPSF}
\includegraphics{HusFig3bbm.EPSF}
\caption{The function $Q_{\lambda \theta}$ for the state $\ket{n}_{10}$, with
$n=6$ and $\theta =0$.  In the top diagram $\lambda = 0.5$.
In the bottom diagram $\lambda = 0.25$.  See Eq.~(\ref{eq:  GenHusFock}).
It can be seen that at $\lambda = 0.25$ the function is beginning to take
its asymptotic form, as given by Eq.~(\ref{eq:  GenHusAsymp}).}
\label{fig:  GenHusLam}
\end{figure}

\begin{figure}
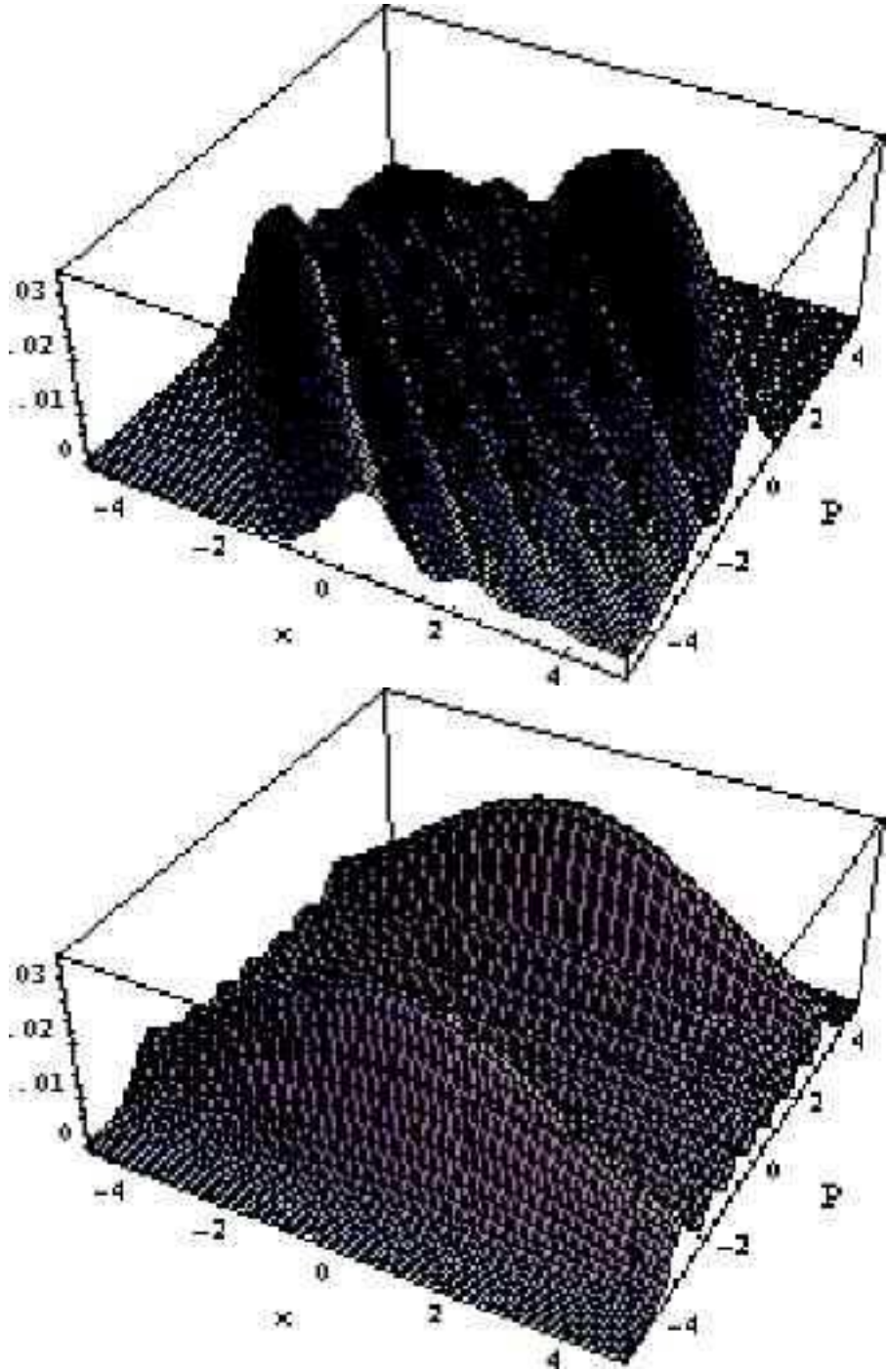

\includegraphics{HusFig4abm.EPSF}
\includegraphics{HusFig4bbm.EPSF}
\caption{The function $Q_{\lambda \theta}$ for the state $\ket{n}_{10}$, with
$n=6$ and $\lambda =0.25$.  In the top diagram $\theta = \frac{\pi}{4}$.
In the bottom diagram $\theta = \frac{\pi}{2}$.  See Eq.~(\ref{eq:  GenHusFock}).
It can be seen that the effect of varying $\theta$ is simply to rotate
the distribution.  This is a consequence of the rotational symmetry of the 
state $\ket{n}_{10}$.}
\label{fig:  GenHusTheta}
\end{figure}

As an example, consider the Fock state $\ket{n}=\ket{n}_{10}$.  The
$Q$-function is
\begin{equation}
  Q(x,p) = Q_{10} (x,p) = \frac{1}{2 \pi n!} E^n e^{-E}
\label{eq:  qfuncFock}
\end{equation}
where $E=\frac{1}{2} (x^2 + p^2)$ is the classical energy.  
Using Eqs.~(\ref{eq:  GenHusInQLess}) and~(\ref{eq:  GenHusInQGreat}) one finds
\begin{align}
 Q_{\lambda \theta} (x, p)
& = \frac{ \sech \eta \, \left|\tanh\eta\right|^n}{2^{n+1} \pi n!}
  \left|
  H_{n} \left( \left| \cosech 2 \eta\right|^{\frac{1}{2}} z_{\lambda \theta} \right) \right|^2
\notag
\\
& \hspace{2.0 in}
    \times
    \exp \left[ - \frac{1}{2} \sech \eta
                    \left(\lambda^{-1} x_{\theta}^{2} + \lambda p_{\theta}^{2} \right)
         \right]
\label{eq:  GenHusFock}
\end{align}
where $\eta = \ln \lambda$, $z_{\lambda \theta}=\frac{1}{\sqrt{2}} (\lambda^{-1} x_{\theta} + i\lambda
p_{\theta})$
and $H_{n}$ is a Hermite polynomial.
$Q_{10}$ is peaked about the classical orbit at $E=n$.  The behaviour of 
$Q_{\lambda\theta}$ for other values of $\lambda$, $\theta$ is illustrated in
Figures~\ref{fig:  GenHusLam} and~\ref{fig:  GenHusTheta}.

The functions $Q_{\lambda \theta}$ are obtained from the Wigner function
by smoothing it in various ways.  Different smoothings extract different features of the information
present in the Wigner function.  No single function extracts all the information
(in explicate form).  If one wants a complete description, 
then it is necessary to consider the complete set of functions
(or else to use the continuation to complex values of $x$ and $p$).

The Wigner function describes the state much more efficiently, using the 
single function $W$, instead
of the two-parameter family of functions $Q_{\lambda \theta}$ (restricted to real
values of their arguments).  
This is an important advantage of the Wigner function. 
However, it can also be a disadvantage  since, precisely because it represents the
information in a very compact form, the Wigner function can be  hard
to interpret.  
The function $Q_{\lambda \theta}$, by contrast, describes the result of making a 
retrodictively optimal determination of phase space location.  As such it has an
immediate physical interpretation, and the picture it presents is much easier to
assimilate intuitively.

The functions $Q_{\lambda \theta}$ can be thought of as a set of
alternative views, or perspectives of a single underlying object.  They might be
compared with a set of photographs of a building taken from various directions. 
The description provided by any single photograph is incomplete. 
In order to get an adequate impression of the building as a whole, one needs to
see how it looks from every direction.

It appears to us that this discussion provides
some additional insight into the intuition underlying Bohr's concept of
complementarity.  In his book
\emph{Interpreting the Quantum World} Bub~\cite{Bub} uses Escher's 
well-known print \emph{Waterfall} as ``a visual metaphor for a quantum world, 
in which the properties of physical systems `fit together' in different
classical or Boolean perspectives that cannot be put together into 
a single Boolean framework''.   The above considerations may be taken
as an alternative illustration of what is, we believe, essentially the 
same point.
Measurements performed for one particular pair of values of
$\lambda$, $\theta$ are not sufficient to exclude the possibility that the observed
probability distribution results from an underlying classical distribution $\Gamma$
\emph{via}  Eq.~(\ref{eq:  ClassDist}).  
It is only when one considers the complete set of functions $Q_{\lambda \theta}$ that
one realises that they cannot all be accounted for in the
same way, in terms of a single classical distribution $\Gamma$ (except in
those cases where the Wigner function is itself
a possible classical distribution~\cite{PosWig}).


\newpage
\newpage
\end{document}